\documentclass{article}

\usepackage{tabularx,booktabs}

\usepackage{PRIMEarxiv}

\usepackage[utf8]{inputenc} % allow utf-8 input
\usepackage[T1]{fontenc}    % use 8-bit T1 fonts
\usepackage{hyperref}       % hyperlinks
\usepackage{url}            % simple URL typesetting
\usepackage{booktabs}       % professional-quality tables
\usepackage{amsfonts}       % blackboard math symbols
\usepackage{nicefrac}       % compact symbols for 1/2, etc.
\usepackage{microtype}      % microtypography
\usepackage{lipsum}
\usepackage{fancyhdr}       % header
\usepackage{graphicx}       % graphics
\graphicspath{{media/}}     % organize your images and other figures under media/ folder
\usepackage{seqsplit}
\usepackage{geometry}

\title{ProtAgents: Protein discovery via large language model multi-agent collaborations combining physics and machine learning 
\thanks{\textit{\underline{Citation}}: 
\textbf{A. Ghafarollahi, M.J. Buehler. arXiv, DOI:000000/11111., 2024}} 
}

\author{
  Alireza Ghafarollahi \\
  Laboratory for Atomistic and Molecular Mechanics (LAMM)\\Massachusetts Institute of Technology\\ 77 Massachusetts Ave.\\ Cambridge, MA 02139, USA\\ \\
   \And
  Markus J. Buehler \\
  Laboratory for Atomistic and Molecular Mechanics (LAMM)  \\
  Center for Computational Science and Engineering\\ Schwarzman College of Computing,\\ Massachusetts Institute of Technology\\77 Massachusetts Ave.\\Cambridge, MA 02139, USA\\ \\
  Correspondence: \texttt{mbuehler@MIT.EDU} \\
}

\begin{document}
\maketitle

\begin{abstract}
Designing \textit{de novo} proteins beyond those found in nature holds significant promise for advancements in both scientific and engineering applications. Current methodologies for protein design often rely on AI-based models, such as surrogate models that address end-to-end problems by linking protein structure to material properties or vice versa. However, these models frequently focus on specific material objectives or structural properties, limiting their flexibility when incorporating out-of-domain knowledge into the design process or comprehensive data analysis is required. In this study, we introduce ProtAgents, a platform for \textit{de novo} protein design based on Large Language Models (LLMs), where multiple AI agents with distinct capabilities collaboratively address complex tasks within a dynamic environment. The versatility in agent development allows for expertise in diverse domains, including knowledge retrieval, protein structure analysis, physics-based simulations, and results analysis. The dynamic collaboration between agents, empowered by LLMs, provides a versatile approach to tackling protein design and analysis problems, as demonstrated through diverse examples in this study. The problems of interest encompass designing new proteins, analyzing protein structures and obtaining new first-principles data -- natural vibrational frequencies -- via physics simulations. The concerted effort of the system allows for powerful automated and synergistic design of \textit{de novo} proteins with targeted mechanical properties. The flexibility in designing the agents, on one hand, and their capacity in autonomous collaboration through the dynamic LLM-based multi-agent environment on the other hand, unleashes great potentials of LLMs in addressing multi-objective materials problems and opens up new avenues for autonomous materials discovery and design.
\end{abstract}

% keywords can be removed
\keywords{Multi-agent modeling \and large language model (LLM) \and physics-inspired machine learning \and protein design}

\section{Introduction}
Proteins, the building blocks of life, serve as the fundamental elements of many biological materials emerging from natural evolution over the span of 300 million years. Protein-base biomaterials like silk, collagen and tissue assemblies such as skin exhibit diverse structural features and showcase unique combinations of material properties. The underlying sequences of amino acids (AAs) in a protein determines its unique there-dimensional structure, which, in turn, dictates its specific biological activity and associated outstanding properties. This inherent relationship has inspired scientists in the field of materials design and optimization to draw valuable insights from nature for creating novel protein-based materials. The diversity in protein design is immense, with over $20^{100}$ possible AA sequences for just a relatively small 100-residue protein. However, the natural evolutionary process has sampled only a fraction of this vast sequence space. This leaves a substantial portion uncharted, presenting a significant opportunity for the \textit{de novo} design of proteins with potentially remarkable properties.\cite{Huang2016} Despite this potential, the extensive design space, coupled with the costs associated with experimental testing, poses formidable challenges in \textit{de novo} protein design. Navigating this intricate landscape necessitates the development of a diverse set of effective tools enabling the targeted design of \textit{de novo} proteins with specific structural features or properties.

Over the past years, in the field of \textit{de novo} protein design, data-driven and machine learning methods have emerged as powerful tools, offering valuable insights and accelerating the discovery of novel proteins with desired properties\cite{Notin2022,Ingraham2019,Wu2022,Anand2022,Eguchi2022,Rives2021,Madani2020,Anand2022b,Greener2018,Riesselman2018,Alley2019,Watson2023,Anishchenko2021,Ingraham2023}. These methods have opened great avenues for predicting structure, properties, and functions of proteins solely based on their underlying AA sequence. For instance, the development of deep learning (DL)-based AlphaFold 2 marked a significant breakthrough in the field of 3D folding protein prediction with a level of accuracy that in some cases rivaled expensive and time-consuming experimental techniques.\cite{Jumper2021}  Moreover, deep learning-based models have been developed to explore structure-property relationships in the analysis and design of proteins. These models encompass a broad spectrum of structural and mechanical properties, serving either as constraints or target values. For example, various DL-models developed predict the secondary structure of proteins from their primary sequences. Prediction of mechanical properties of spider silk protein sequences have been enabled by DL models\cite{Yu2022,Elnaggar2022,Mirabello2013,Pollastri2002,Zhang2018,Pollastri2005}. Moreover, DL-based models such as graph neural networks\cite{Guo2022} and transformer-based language models\cite{Hu2022} show enhanced accuracy in predicting the protein natural frequencies compared to physics-based all-atom molecular simulations. The development of such DL models significantly reduces the cost of screening the vast sequence space to target proteins with improved or optimized mechanical performance.
In the field of \textit{de novo} protein design, data-driven and machine learning methods have emerged as powerful tools, offering valuable insights and accelerating the discovery of novel proteins with desired properties\cite{Notin2022,Ingraham2019,Wu2022,Anand2022,Eguchi2022,Rives2021,Madani2020,Anand2022b,Greener2018,Riesselman2018,Alley2019,Watson2023,Anishchenko2021,Ingraham2023}. These methods have opened great avenues for predicting structure, properties, and functions of proteins solely based on their underlying AA sequence. For instance, the development of deep learning (DL)-based AlphaFold 2 marked a significant breakthrough in the field of 3D folding protein prediction with a level of accuracy that in some cases rivaled expensive and time-consuming experimental techniques.\cite{Jumper2021}  Moreover, deep learning-based models have been developed to explore structure-property relationships in the analysis and design of proteins. These models encompass a broad spectrum of structural and mechanical properties, serving either as constraints or target values. For example, various DL-models developed predict the secondary structure of proteins from their primary sequences. Prediction of mechanical properties of spider silk protein sequences have been enabled by DL models\cite{Yu2022,Elnaggar2022,Mirabello2013,Pollastri2002,Zhang2018,Pollastri2005}. Moreover, DL-based models such as graph neural networks\cite{Guo2022} and transformer-based language models\cite{Hu2022} show enhanced accuracy in predicting the protein natural frequencies compared to physics-based all atomistic simulations. The development of such DL models significantly reduces the cost of screening the vast sequence space to target proteins with improved or optimized mechanical performance. 

\begin{figure*}
 \centering
 \includegraphics[width=\textwidth]{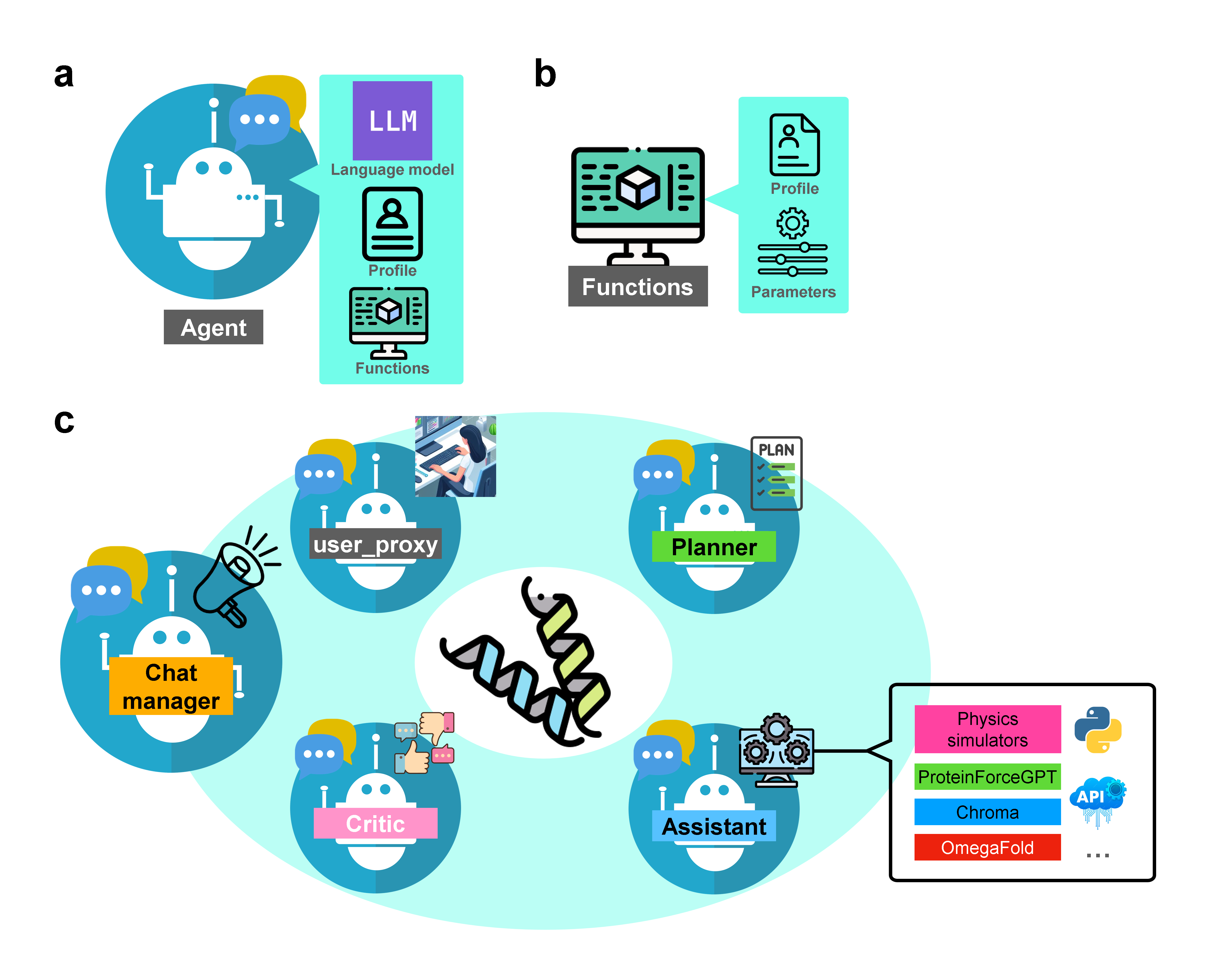}
 \caption{\textbf{Multi-agent AI framework for automating protein discovery and analysis.} \textbf{a}, A genetic agent structure in a multi-agent modeling environment that can communicate via language, has a focus defined by a profile, and has access to custom functions. \textbf{b}, A function is customized by a profile and a set of parameters. \textbf{c}, The structure of a team of agents, each with special expertise, that communicate to each other and allow for mutual correction and a division of labor. Given different profiles for each agent, agents are designed that are expert on describing the problem (\textit{user\_proxy}), plan making (\textit{planner}), function executing (\textit{assistant}), and result evaluation (\textit{critic}). The whole process is automated via a dynamic group chat under the leading chat manager, offering a versatile approach in solving challenging tasks in the context of protein design and analysis without human intervention.}
 \label{fig:agents}
\end{figure*}

A frontier, however, that still exists is how we can create intelligent tools that can solve complex tasks and draw upon a diverse set of knowledge, tools and abilities. Another critical issue is that the combination of purely data-driven tools with physics-based modeling is important for accurate predictions. Moreover, such tools should ideally also be able to retrieve knowledge from, for instance, the literature or the internet. All these aspects must be combined in a nonlinear manner where multiple dependent steps in the iteration towards and answer are necessary to ultimately provide the solution to a task. As we will discuss in this study, such an integration of tools, methods, logic, reasoning and iterative solution can be implemented through the deployment of a multi-agent system driven by sophisticated Large Language Models (LLMs). 

LLMs\cite{Wei2022, Chang2023} have represented a paradigm shift in modeling problems across a spectrum of scientific and engineering domains\cite{Jablonka2023,Buehler2023MeLM,Buehler2023MechGPT,Boyko2023,Madani2020,Ni2023Agent,Brodnik2023,Tinn2023,Hu2023,Buehler2023ontologic,Luu2023,Peng2021,Brown2020,Bates1995,Thoppilan2022,Chowdhery2023}. Such models, built upon attention mechanism and transformer architectures\cite{Vaswani2017}, have emerged as powerful tools recently in the field of materials science and related areas, contributing to various aspects ranging from knowledge retrieval to modeling, design, and analysis.  For example, models such as ChatGPT and the underlying GPT-4 architecture\cite{OpenAI2023}, part of the Generative Pretrained Transformer (GPT) class, demonstrate exceptional proficiency in mastering human language, coding\cite{Chen2021}, logic and reasoning\cite{Bubeck2023SparksGPT-4b}. Recent studies highlight their ability to proficiently program numerical algorithms and troubleshoot code errors across several programming languages like Python, MATLAB, Julia, C, and C++\cite{austin2021program}. The GPT class of LLMs has also represented a new paradigm in simulating and predicting the materials behavior under different conditions\cite{Buehler2023MeLM}, a field of materials science often reserved for conventional deep learning frameworks\cite{Guo2021} such as Convolutional Neural Networks\cite{Li2022,Keiron2015}, Generative Adversarial Networks\cite{Buehler2022multimaterial,Yang2021,Repecka2021}, Recurrent Neural Networks22,54,55\cite{Pollastri2002,Cao2017,Hsu2020},and Graph Neural Networks\cite{Guo2022, Lu2022,Strokach2020,You2021,Yang2022}. Moreover, due to their proficiency in processing and comprehending vast amount of different types of multimodal data, LLMs show promising capabilities in materials analysis and prediction application including key knowledge retrieval\cite{Buehler2023ontologic}, general language tasks, hypothesis generation\cite{Buehler2023MechGPT}, and structure-to-property mapping\cite{Buehler2023MeLM,Lu2023kaplan}.     

\begin{figure*}
 \centering
 \includegraphics[scale=0.48]{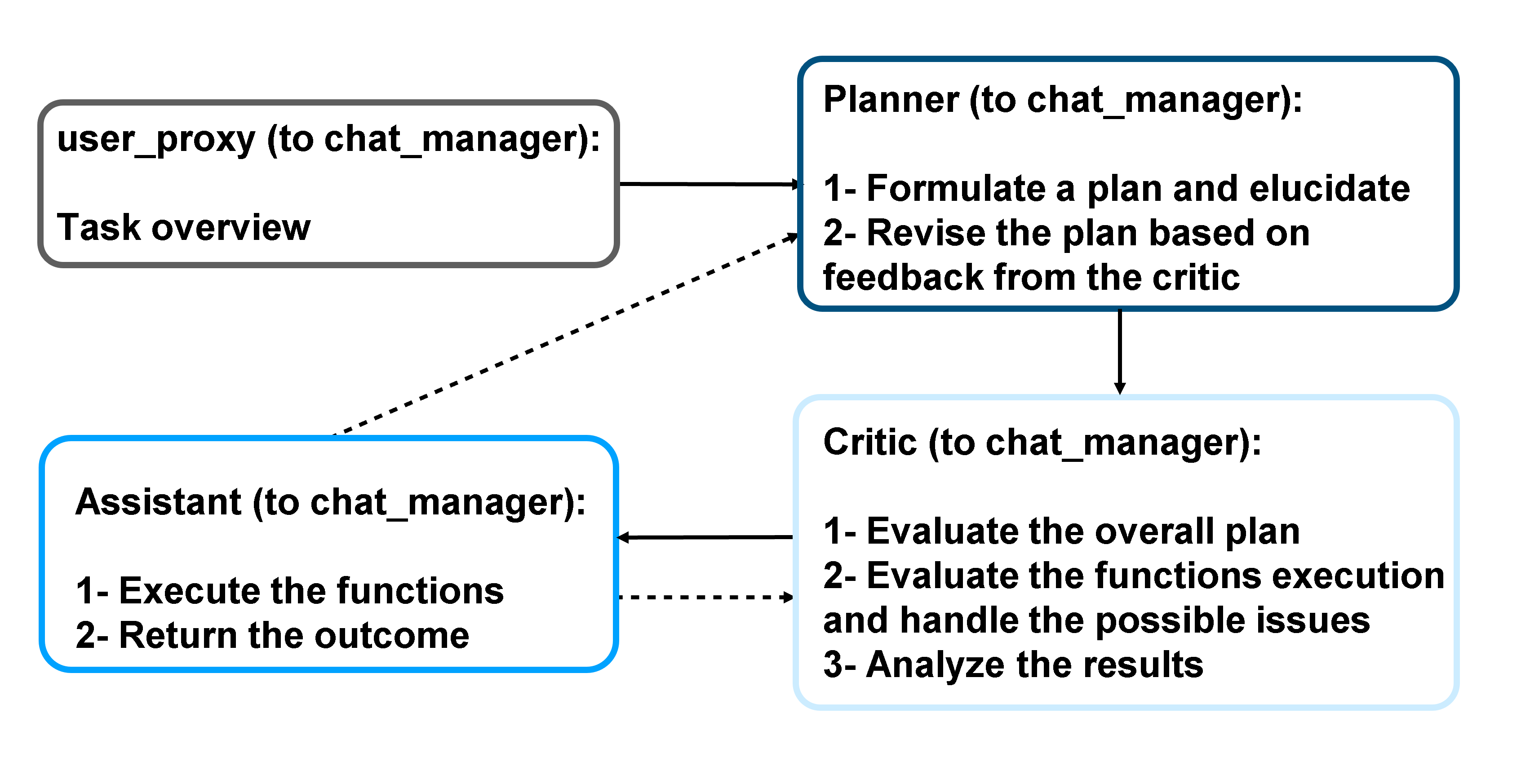}
 \caption{\textbf{A generic flowchart showing the dynamic interaction between the multi-agent team members organized by the group chat manager to solve protein design and analysis problems.} The manager selects the working agents to collaborate in the team work based on the current context of the chat, thus forming close interactions and enabling mutual corrections.}
 \label{fig:generic}
\end{figure*}

At the same time, LLMs are typically not best equipped to solve specific physics-based forward and inverse design tasks, and are often focused on leveraging their conversational capabilities. Here, LLMs have been instrumental in powering conversable AI agents, facilitating the transition from AI-human conversations to AI-AI or AI-tools interactions for increased autonomy.\cite{Ni2023Agent,Buehler2023ontologic,Zhang2023Building,Wang2023,Xi2023}  This capability represents a significant advancement, enabling intelligent mediation, fostering interdisciplinary collaboration, and driving innovation across disparate domains, including materials analysis, design, and manufacturing. The overall process could be deemed as adapting a problem-solving strategy dictated and directed by the AI system comprised of different agents. Thereby, the entire process can be AI automated with reduced or little human intervention. Depending on the complexity of the problem, using the idea of labor division, the agents have the capability to break the overall task into subtasks for which different agents or tools are used consecutively to iteratively solve the problem until all subtasks have accomplished and the solution has achieved. There is no intrinsic limitation in defining the type of tools, making the multi-agent model a versatile approach in addressing problems across scales and disciplines. The tools could range from a simple linear mathematical function to sophisticated deep neural network architectures. The multi-agent strategy has been explored in materials and mechanics applications through in earlier work\cite{Buehler2023MechGPT} and was further explored in the context of molecular modeling tasks\cite{Buehler2023ontologic}.  

In this paper, we propose a multi-agent strategy to the protein design problems by introducing ProtAgents, a multi-agent modeling framework to solve protein-related analysis and design problems by leveraging customized functions across domains and disciplines. The core underpinning concept of the multi-agent systems is the use state-of-the-art LLMs combined with a series of other tools. The LLM backbone demonstrate exceptional abilities in analysis, rational thinking, and strategic planning, essential for complex problem-solving. Leveraged by these capabilities, the proposed model aims to reduce the need for human intervention and intelligence at different stages of protein design. The agent model consists a suite of AI and physics based components such as: 

\begin{itemize}
    \item Physics simulators: obtain new physical data from simulations, specifically normal modes and vibrational properties by solving partial differential equations (PDEs)  
    \item Generative AI model: conditional/unconditional \textit{de novo} protein design, based on a denoising diffusion model 
    \item Fine-tuned transformer model: predict mechanical properties of proteins from their sequence
    \item Retrieval agent: retrieve new data from a knowledge database of scientific literature 
\end{itemize}
The resulting model has the ability to go beyond the conventional DL models by integrating new physical data or information across disciplines, for instance via writing and executing code that solves differential equations or other physics-based numerical methods, or that conducts retrieval augmented generation (RAG)\cite{Lewis2020}. A tool-baked agent has access to various tools and functions with different functionalities that may be called upon, for instance, to predict a specific protein property or to obtain new physical data such as natural frequency from physics-based simulations. The versatility of the approach in solving complex tasks is exhibited by providing a series of experiments in the context of proteins design, modeling, and data analysis. 

The plan of this paper is as follows. In Section \ref{sec:results}, we present an overview of the multi-agent framework developed to tackle multi-objective complex tasks. Subsequently, we delve into a series of experiments where each task is initially introduced, followed by a detailed examination of various aspects throughout the problem-solving process by the multi-agent teamwork. A comprehensive discussion regarding the multi-agent framework and future prospects is provided in Section \ref{sec:conclusions}.

\section{Results and Discussion}\label{sec:results}
We present a series of computational experiments aimed at evaluating the effectiveness and potential of a multi-agent modeling framework for various challenges within the domain of protein modeling, design, and analysis. The multi-agent framework consists of a team of agents, each powered by a state-of-the-art general purpose large language model, GPT-4,\cite{OpenAI2023} accessed via the OpenAI API\cite{OpenAI_API} and characterized by a unique profile that details its role, and communication protocols, such as sharing information and engaging with humans via language as shown in Figure \ref{fig:agents}a.  Furthermore, agents are given access to a set of tools with various functionalities across domains. As shown in Figure \ref{fig:agents}b each function is characterized by a descriptive profile and input parameters. The outline of the proposed multi-agent framework is shown in Figure \ref{fig:agents}c, illustrating the collaborative efforts of a team of agents with the following entities

\begin{figure*}
 \centering
 \includegraphics[width=\textwidth]{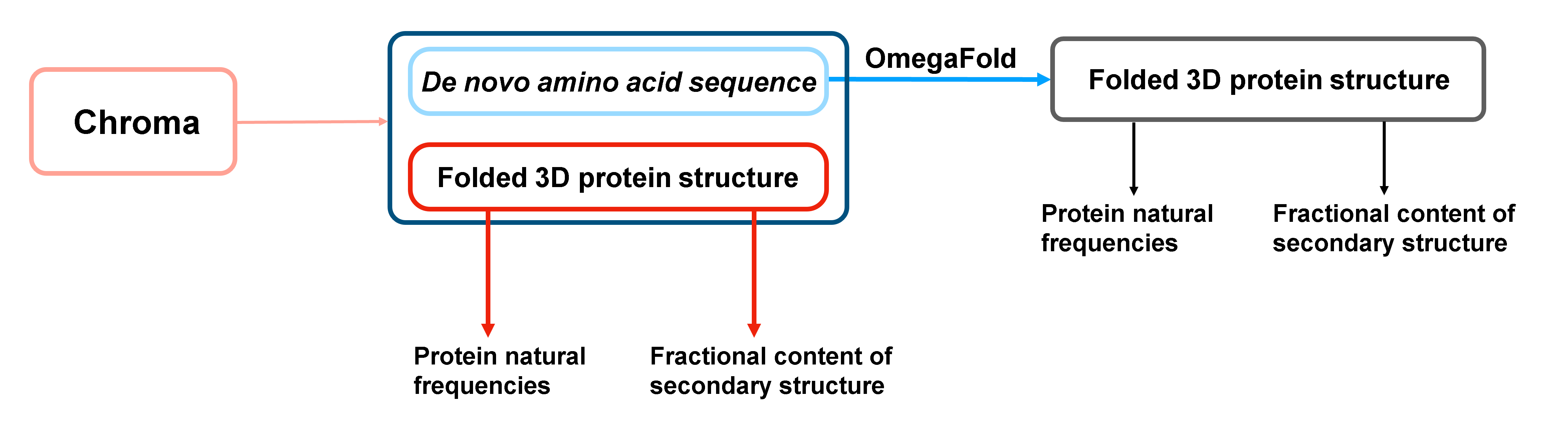}
 \caption{\textbf{Overview of the multi-agent work to solve the complex task posed in experiment II, Section \ref{sec:exp_2}}. First the multi-agent uses Chroma to generate \textit{de novo} protein sequences and then computes natural frequencies and secondary structures content for the generated structures. Next, from \textit{de novo} AA sequences, the model finds the 3D  folded structures using OmegaFold and finally computes the frequencies and secondary structure content for the protein structures.  The results obtained from the Chroma and OmegaFold 3D protein structures are compared in Figure~\ref{fig:chroma_fold}.  }
 \label{fig:exp2}
\end{figure*}

\begin{itemize}
\item “User”: human that poses the question
\item 	“Planner”: develops a plan to solve the task. Also suggests the functions to be executed.
\item 	“Assistant”: who has access to all the customized functions, methods, and APIs and executes them to find or compute the relevant data necessary to solve the task
\item 	“Critic”: Responsible for providing feedback about the plan developed by “planner” as well as analyzing the results and handling the possible mistakes and providing the output to the user.
\end{itemize}
The agents are organized into a team structure, overseen by a manager who coordinates overall communication among the agents. A generic structure showing the dynamic collaboration between the team of agents proposed in the current study is depicted in Figure \ref{fig:generic}. Moreover, Table \ref{tbl:agents_profile} lists the full profile for the agents recruited in our multi-agent framework. Further details can be found in the \textbf{Materials and Methods} section \ref{sec:methods}.

\begin{table*}
\tiny
\centering
\caption{\textbf{The profiles of the agents implemented in the current study to solve multi-objective tasks in the context of protein design and analysis.}}
  \label{tbl:agents_profile}
\begin{tabularx}{\textwidth}{llX}
\toprule
\textbf{Agent \# } & \textbf{Agent role} & \textbf{Agent profile} \tabularnewline
\midrule
1 & user\_proxy &  user\_proxy. Plan execution needs to be approved by user\_proxy \tabularnewline
\midrule
2 & Planner & Planner. You develop a plan. Begin by explaining the plan. Revise the plan based on feedback from the critic and user\_proxy, until user\_proxy approval. 
The plan may involve calling custom function for retrieving knowledge, designing proteins, and computing and analyzing protein properties. You include the function names in the plan and the necessary parameters.
If the plan involves retrieving knowledge, retain all the key points of the query asked by the user for the input message.\tabularnewline
\midrule
3 & Assistant &  Assistant. You have access to all the custom functions. You focus on executing the functions suggested by the planner or the critic. You also have the ability to prepare the required input parameters for the functions. \tabularnewline
\midrule
4 & Critic &  user\_proxy. You double-check the plan, especially the functions and function parameters. Check whether the plan included all the necessary parameters for the suggested function. You provide feedback.	 \tabularnewline
\midrule
5 & Group chat manager &  You repeat the following steps: Dynamically selecting a speaker, collecting responses, and broadcasting the message to the group. \tabularnewline
\bottomrule
\end{tabularx}
\end{table*}

It is noteworthy that critical issues in the realm of protein design surpass the capabilities of mere Python code writing and execution. Instead, addressing these challenges necessitates the utilization of external tools specifically tailored for protein design and analysis, and the writing, adaptation, correction and execution of code depends nonlinearly on the progression of the solution strategy that is developed by the system. 

The tools are incorporated into the model via the Assistant agent who oversees executing the tools. To assess the performance of the multi-agent framework in handling complex interdisciplinary tasks, we have defined a rich library of functions each with special powers in solving the protein problems. Each function has a distinct profile describing its functionally and takes one or more required entities as the input. The functions provide the ability to, for instance, retrieve knowledge, perform protein folding, analyze the secondary structure, and predict some parameters through a pre-trained autoregressive language model. Additionally, a function can carry out simulations to compute the protein natural frequencies, thus allowing the model to integrate the new physics-based data. A full list of functions implemented in the current study is provided in \textbf{Table S1} in the supplementary material. 

Given the complexities residing in the protein design problems, the primary contribution of our multi-agent framework lies in assessing whether the team of agents can discern the requisite tools for a given query and evaluating the framework's capability to initiate the execution of these tools, along with providing the necessary inputs. The designated tasks are intentionally designed to be sufficiently complex, involving multiple subtasks where, for some cases, the execution of each depends on the successful completion of the preceding ones. This design showcases the model’s capacity for the automated handling of intricate tasks, eliminating or substantially reducing the need for human intervention. Although the multi-agent framework allows for the human intervention at different stages, we skip that to further examine the team’s capability in handling different possible situations, for instance in case of a failure. 

\subsection{Experiment I: Knowledge retrieval, computations, and analysis}\label{sec:exp_1}
As the first example, we pose a task which involves a multi-step knowledge retrieval, analysis, and computations for a set of protein PDB identifiers. We start the first round of conversation by posing the following question:

\begin{quote}
    \textbf{user\_proxy (to chat\_manager):}
    
        \textit{Can you provide examples of protein names whose mechanical properties have been studied through experiments?}
    
\end{quote}
The planer then correctly suggests the function “retrieve\_content” to be executed with the argument “examples of protein names whose mechanical properties have been studied through experiments”. Upon execution of the function, the assistant provides us with a list of protein names. Upon inspection, we find that the agent has successfully identified experimentally studied proteins, despite an abundance of information on proteins studied theoretically, for instance, through coarse-grain simulations. Since we are interested in the PDB ids, we continue the chat by a follow-up question “Can you provide me with the PDB ids for these proteins?” when “user\_proxy” is being asked to provide feedback to chat manager. Again, the planner suggests “retrieve\_content” function with the following message

\begin{quote}
    \textbf{Planner (to chat\_manager):\\}
    
        \textit{***** Suggested function Call: retrieve\_content *****
Arguments: 
\{
  "message": "PDB ids for I27 domain from titin, fibronectin type III domains, E2lip3, T4 lysozyme, ubiquitin, titin immunoglobulin domains, cell adhesion molecule, spectrin repeats, and tenascin"
\}
}
\end{quote}
The “\textit{Assistant}” agent then calls the function and gives the following output:

\begin{quote}
    \textbf{Assistant (to chat\_manager):\\}
    
    \textit{Here are the PDB IDs for the proteins you mentioned:\\
\\
    1. I27 domain from titin: 1wit\\
    2. Ubiquitin: 1ubq\\
    3. T4 lysozyme: 1nct\\
    4. Titin immunoglobulin domains: 1tit, 1qjo\\
    5. E2lip3: 1hz6/2ptl\\
    6. Tenascin: 1ten\\
    7. Spectrin repeats: 1emb\\
    8. FnIII: 1g1c, 1fnh, 1vsc, 1cfc, 1fnf/1ttf/1ttg\\
    9. Cell adhesion molecule: 1bni/1bnr, 1b6i, 1rsy/1dqv, 1aj3, 1ksr/1whl, 1u4q, 1hci, 1n11
}
    
\end{quote}
Upon careful examination of the results, we observe that, despite all the PDB ids exist in the source database, the PDB ids do not quite match with the protein names except for a few cases. Nevertheless, note that the error is caused by the poor performance of the “\textit{retrieve\_content}” function and the team of agents cannot mitigate that as they have no access to the knowledge database.  As such, we continue to test the agent-team capability in more challenging queries centered around computational tasks and physics-based simulations by assigning the following task in the next round of conversation

\begin{quote}
    \textbf{user\_proxy (to chat\_manager):\\}
    
        \textit{For the first 10 protein PDB ids in the list, if the AA sequence length of the protein is less than 128, do the following: (a) analyze the secondary structure of the protein, (b) compute its first 13 natural frequencies, (c) determine its structure classification.}
    
\end{quote}
The above is a complex multi-step analysis and computation task encompasses aspects such as secondary structures analysis, natural frequency calculations, and structure classification. Additionally, the task is subject to an initial condition that must be satisfied before proceeding through the next sequence of steps, adding an extra layer of complexity. In response, the planner comes up with a detailed plan which consists of all the actions that need to be taken to complete the task. Moreover, the plan mentions all the corresponding functions that need to be executed to accomplish the task. More importantly, the “\textit{planner}” perfectly realizes to fetch the protein structures first before starting to analyze the secondary structure, although it was not explicitly mentioned in the task query. 

The teamwork proceeds by a follow-up feedback provided by the “\textit{critic}” agent about all the plan steps and functions which is concluded by the following statement

\begin{quote}
    \textbf{critic (to chat\_manager):\\}
    
        \textit{Overall, the plan rightly identifies all the necessary functions and respective parameters for each step. It illustrates a good understanding of what needs to be done to fulfill the user request. Good job!}
    
\end{quote}
Therefore, the positive feedback from the “\textit{critic}” further supports the good performance of the planner in addressing all the critical steps required to accomplish the tasks. 

The “\textit{assistant}” agent then follows the plan by calling and executing the corresponding functions, starting with AA length calculation, until all the steps have been undertaken. The results show that all the inputs to the functions are properly identified and provided and the functions are executed without any error. The conditional statement included in the tasks is also correctly satisfied for each protein, that is the computations are conducted only if the sequence length is less than 128 and are emitted otherwise. For instance, for the protein with PDB id “\textit{1hz6}” the AA length is returned as 216 by the “\textit{assistant}” which is then followed by the following message from the “\textit{critic}” 

\begin{quote}
    \textbf{critic (to chat\_manager):\\}
    
        \textit{The length of the amino-acid sequence for the PDB id '1hz6' is 216, which is greater than 128. Therefore, we will not perform the steps (a) analyze the secondary structure of the protein, (b) compute its first 13 natural frequencies, and (c) determine its structure classification for the '1hz6'. \\
        \\
        Moving to the next PDB id in the list...}
    
\end{quote}

After completion of all the tasks, the assistant returns a summary of all the results for each protein as representatively shown below for PDB id “\textit{1wit}”:

\begin{quote}
    \textbf{assistant (to chat\_manager):\\}
    
        \textit{For PDB id '1wit':\\
- It has an amino acid sequence length of 93.\\
- The analysis of its secondary structure yielded the following composition: H: 0\%, B: 3.23\%, E: 51.61\%, G: 3.23\%, I: 0\%, T: 13.98\%, S: 5.38\%, P: 0\%, -: 22.58\%.\\
- Its first 13 natural frequencies are [4.3755, 5.0866, 5.5052, 6.7967, 7.908, 8.1947, 9.0166, 9.8528, 11.0632, 11.3968, 11.7355, 12.1279, 12.3498].\\
- Its CATH classification is 2.60.40.10.\\
}
    
\end{quote}
The results suggest that the framework effectively retains all outputs, demonstrating its strong memory even in the face of diverse and extended results. As the last round of conversation, we ask to save all the results which allows us to load them at later time for other purposes:
\begin{quote}
    \textbf{user\_proxy (to chat\_manager):\\}
    
        \textit{Save all the results into a csv file where each row corresponds to a protein and each column represent     a protein property, utilizing proper identifiers.}
    
\end{quote}
In response, the planner suggests to call the python function “\textit{save\_to\_csv\_file}”. The main task here is to generate the dictionary of results in JSON and with appropriate structure as instructed by the user. However, we see that upon generating the JSON data and inputting it into the function by the "\textit{assistant}" agent, the following error occurs:

\begin{quote}
    \textit{***** Response from calling function "save\_to\_csv\_file" *****\\
Error: Expecting ',' delimiter: line 1 column 141 (char 140)
}
\end{quote}
Without any human intervention, the agent team is able to resolve the issue by mutual correction. In particular, the “\textit{critic}” identifies the cause of error by writing

\begin{quote}
    \textbf{Critic (to chat\_manager):\\}
    
        \textit{ It seems like there is an error in the JSON format being used to feed data into the csv conversion function. The JSON string appears to be improperly formed leading the function to fail. The values for "Secondary Structure" and "First 13 Frequencies" are stringified JSON objects themselves, which maybe causing}
    
\end{quote}
The “\textit{assistant}” then reconstructs the JSON file from the output results and is able to successfully execute the function and thus save the results in a csv file as shown in \textbf{Table \ref{tbl:exp1}}. The complete group chat records can be found in \textbf{Table S2} of the Supplementary material.

\begin{table*}
\tiny
\centering
\caption{\textbf{The results generated by the group chat in the form of a CSV file, without human intervention, for the experiment I, Section \ref{sec:exp_1}.}}
  \label{tbl:exp1}
\begin{tabularx}{\textwidth}{llXXl}
\toprule
\textbf{Protein ID\ \# } & \textbf{Amino Acid Length} & \textbf{Secondary Structure} & \textbf{First 13 Frequencies} & \textbf{CATH Classification} \tabularnewline
\midrule
1wit & 93 & [{'H': 0.0, 'B': 3.23, 'E': 51.61, 'G': 3.23, 'I': 0.0, 'T': 13.98, 'S': 5.38, 'P': 0.0, '-': 22.58}] & [4.3755, 5.0866, 5.5052, 6.7967, 7.908, 8.1947, 9.0166, 9.8528, 11.0632, 11.3968, 11.7355, 12.1279, 12.3498] & 2.60.40.10
\tabularnewline
\midrule
1ubq & 76 & [{'H': 15.79, 'B': 2.63, 'E': 31.58, 'G': 7.89, 'I': 0.0, 'T': 15.79, 'S': 5.26, 'P': 5.26, '-': 15.79}] & [0.7722, 1.0376, 1.5225, 1.6534, 2.5441, 2.9513, 3.2873, 3.7214, 4.1792, 4.3437, 4.3908, 4.6551, 5.1631] & 3.10.20.90
\tabularnewline
\midrule
1nct & 106 & [{'H': 0.0, 'B': 4.08, 'E': 35.71, 'G': 0.0, 'I': 0.0, 'T': 2.04, 'S': 21.43, 'P': 0.0, '-': 36.73}] & [3.6644, 4.425, 6.5351, 6.7432, 7.1409, 7.1986, 9.0207, 9.2223, 10.3163, 10.7313, 11.5299, 11.6373, 12.5606] & 2.60.40.10
\tabularnewline
\midrule
1tit & 98 & [{'H': 0.0, 'B': 1.12, 'E': 35.96, 'G': 0.0, 'I': 0.0, 'T': 6.74, 'S': 17.98, 'P': 0.0, '-': 38.2}]	&[5.5288, 5.9092, 8.2775, 8.6267, 9.3391, 9.8783, 10.1607, 11.451, 11.5896, 11.7052, 12.1498, 12.6082, 13.8622]	& 2.60.40.10
\tabularnewline
\midrule
1qjo & 80 & [{'H': 0.0, 'B': 2.5, 'E': 40.0, 'G': 0.0, 'I': 0.0, 'T': 8.75, 'S': 13.75, 'P': 0.0, '-': 35.0}]&	[3.8578, 4.4398, 5.4886, 5.7815, 6.6332, 6.9269, 7.2329, 7.6453, 8.2545, 8.3076, 8.6118, 8.7135, 8.8546]	& 2.40.50.100
\tabularnewline
\midrule
2ptl & 78 & [{'H': 15.38, 'B': 1.28, 'E': 30.77, 'G': 0.0, 'I': 0.0, 'T': 7.69, 'S': 19.23, 'P': 0.0, '-': 25.64}]	&[0.0386, 0.1161, 0.2502, 0.5921, 1.1515, 1.5257, 2.0924, 2.6793, 3.4292, 3.9289, 4.2172, 4.6878, 4.8022] &	3.10.20.10
\tabularnewline
\bottomrule
\end{tabularx}
\end{table*}

\subsection{Experiment II: \textit{De novo} protein design using Chroma}\label{sec:exp_2}

An important characteristic of the multi-agent model is its capability in handling very complex tasks in the context of protein design and analysis. This partly stems from the possibility of incorporating customized functions in the model for various purposes from knowledge retrieval to performing physics-based simulations. In this example, we explore the model’s performance in using the state-of-the art models in \textit{de novo} protein design, Chroma\cite{Ingraham2023}, and protein folding, OmegaFold\cite{Wu2022}. Chroma offers a unified approach for protein design that can directly sample novel protein structures and sequences with the possibility to direct the generative design towards the desired functions and properties by appropriate conditioning. OmegaFold is a state-of-the-art folding method without the need of using multiple sequence alignments (MSA) to predict the three-dimensional folded protein structure given its AA sequence. 

In this experiment, we formulate a complex multi-step task with the objective of comparing the two models based on various structural and physical features derived from the folded structures obtained through Chroma and OmegaFold2. We pose the following task through the “\textit{user\_proxy}” agent:

\begin{quote}
    \textbf{user\_proxy (to chat\_manager):\\}
    
        \textit{Design three proteins with sequence length of 120 and analyze their secondary structure and first 10 frequencies. Then fold the proteins and repeat the analysis. Save all the results (including the AA sequences) into a csv file where each row corresponds to a protein and each column represents a protein property, utilizing proper identifiers.}
    
\end{quote}

The “\textit{planner}” then suggests the following plan 

\begin{figure*}
 \centering
 \includegraphics[width=\textwidth]{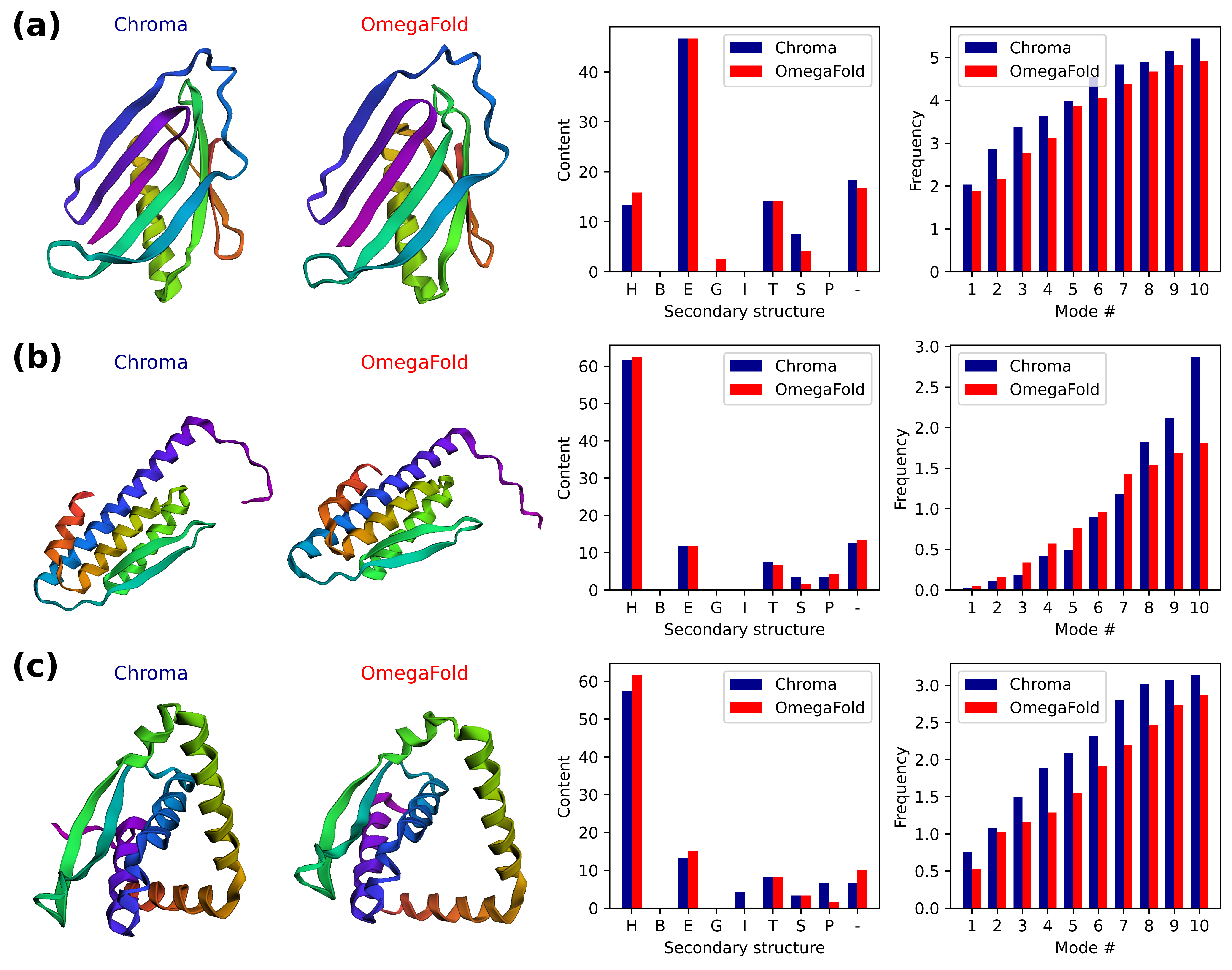}
 \caption{\textbf{The results generated by the multi-agent collaboration for the experiment II, Section \ref{sec:exp_2}.} The first and second columns depict the 3D folded structures of proteins generated by Chroma and OmegaFold2, respectively, while the third and fourth columns represent the fractional content of secondary structures, and first ten natural frequencies for the generated proteins.}
 \label{fig:chroma_fold}
\end{figure*}

\begin{quote}
    \textbf{Planner (to chat\_manager):\\}
        \textit{Plan:
1. Use the `design\_protein\_from\_length` function three times to design three proteins each with a sequence length of 120. I will use 1, 2, 3 as names for the proteins. Save the amino acid sequences and PDB filenames.\\
2. Use the `analyze\_protein\_structure` function to analyze the secondary structure of the proteins. Store the results.\\
3. Use the `calc\_protein\_ANM` function to calculate the first 10 frequencies of the proteins. Store the results.
4. Use the `fold\_protein` function to fold the proteins. \\
5. Repeat steps 2 and 3 with the folded proteins.\\
6. Use the `save\_to\_csv\_file` function to save all results into a csv file where each row corresponds to a protein and each column represents a protein property.\\
7. The identifiers will be named as follows:\\
   - Protein Name\\
   - Amino Acid Sequence (Pre-Fold)\\
   - Secondary Structure (Pre-Fold)\\
   - Frequencies (Pre-Fold)\\
   - Amino Acid Sequence (Post-Fold)\\
   - Secondary Structure (Post-Fold)\\
   - Frequencies (Post-Fold)\\
}
\end{quote}
At the first glance, the plan seems to cover all the details necessary to accomplish the tasks included in the problem statement. However, the “\textit{critic}” agent who is responsible in giving feedback about the plan spots a minuscule error in the saving part of the plan as follows:

\begin{quote}
    \textbf{Critic (to chat\_manager):\\}

    \textit{
    Your plan is almost perfect. However, when a protein is folded, the protein sequence doesn't change, only the structure does, which then affects the secondary structure and possible frequencies you can calculate. So you need to only save Amino Acid Sequence once after designing the proteins. Each protein should have the following column properties:
    }
\\
\\
\textit{
    - Protein Number\\
  - Amino Acid Sequence\\
  - Secondary Structure (Pre-Fold)\\
  - Frequencies (Pre-Fold)\\
  - Secondary Structure (Post-Fold)\\
  - Frequencies (Post-Fold)}\\
\\
\textit{Other than that, your plan works perfectly. So steps will be as follows:}\\
...
\end{quote} 
The correction made by the “\textit{critic}”concerning the sequence length underscores its notable proficiency in comprehending how diverse functions and parameters influence various aspects within the realm of protein design. 
\\
The “\textit{user\_proxy}” agent is then asked to confirm the plan. The “\textit{assistant}” then takes the stage and starts following the plan by calling and executing the functions until all the steps have been undertaken. An overview of the work performed by the “\textit{assistant}” is depicted in Figure \ref{fig:exp2}.  At the end of the computations, the results are formatted into a JSON dictionary to fed into the “save\_to\_csv\_file” function. However, an error related to the JSON dictionary format occurs when executing the function as follows:

\begin{quote}
    \textit{***** Response from calling function "save\_to\_csv\_file" *****\\
Error: Expecting property name enclosed in double quotes: line 2 column 5 (char 6)\\
*************************************************************
}
\end{quote}
The “\textit{critic}” then steps in by making the following comment and suggesting a plan to fix the error as follows:

\begin{quote}
    \textbf{Critic (to chat\_manager):\\}
\textit{I apologize for the mistake. It seems there was an issue with the formatting of the input JSON dictionary for saving to the CSV file. Let's try again with properly formatted JSON.}
\\
\\
\textit{Plan to fix the error:}
\\
\\
\textit{1. Correctly format JSON with double quotes instead of single quotations.}\\
\textit{2. Call 'save\_to\_csv\_file' function again with correctly formatted JSON dictionary.}
\\
\\
\textit{I'll directly proceed with this plan as the error was due to syntax not plan. }

\end{quote}
The critic makes the necessary corrections and suggest the corrected JSON dictionary for the “assistant” to execute the “save\_to\_csv\_file” function. This time, the function is successfully executed and the results are saved into a csv file as shown in \textbf{Table \ref{tbl:exp2}}. At last, the “\textit{critic}” gives an evolution about the whole process:

\begin{quote}
    \textbf{Critic (to chat\_manager):\\}
    
    \textit{The plan was executed successfully and all data including Amino Acid Sequence, Secondary Structure (Pre-Fold and Post-Fold), and Frequencies (Pre-Fold and Post-Fold) for three proteins have been saved into a CSV file named 'protein\_data.csv'. Each row of the file corresponds to a protein and each column represents a protein property.}
\end{quote}    
This experiment showcases a good example of multi-agent collaboration in solving complex tasks in the context of \textit{de novo} protein design and analysis. It specially shows the great capability of the “\textit{critic}” agent in providing valuable feedback to other working agents at different stages of the problem solving endeavor, further assisting the team of agents in handling possible errors without the need for human involvement. Figure \ref{fig:chroma_fold} shows the plots of the generated results including the 3D folded structures. The full conversations can be found in \textbf{Table S3} of the Supplementary material.

\begin{table*}
\tiny
\centering
\caption{\textbf{The final results generated by the group chat in the form of a CSV file, without human intervention, for the second experiment II, Section \ref{sec:exp_2}.} }
  \label{tbl:exp2}
\begin{tabularx}{\textwidth}{lXXXXX}
\toprule
\textbf{Protein Number\ \# } & \textbf{Amino Acid Sequence} & \textbf{Secondary Structure (Pre-Fold)} & \textbf{Frequencies (Pre-Fold)} & \textbf{Secondary Structure (Post-Fold)} &\textbf{Frequencies (Post-Fold)} \tabularnewline
\midrule
1 & \seqsplit{MIIINIKTENGLSITYNSDEKKLELKYTPVKSPEDFKFPEDAKATISEVEYKGKKV IKIDAKLYVSPDLSKAKLTIEVNADISQEEADKIIDEFIKLLESLGNIKLKVTKD GNKYTIEVE} &	{'H': 13.3333333333, 'B': 0.0, 'E': 46.6666666666, 'G': 0.0, 'I': 0.0, 'T': 14.1666666666, 'S': 7.5, 'P': 0.0, '-': 18.33333333333} &	[2.0337, 2.8678, 3.3843, 3.6263, 3.9904, 4.5381, 4.8373, 4.8956, 5.1492, 5.4416] & {'H': 15.8333333333, 'B': 0.0, 'E': 46.666666666, 'G': 2.5, 'I': 0.0, 'T': 14.1666666666, 'S': 4.1666666666, 'P': 0.0, '-': 16.666666666} &	[1.8739, 2.1563, 2.7611, 3.1086, 3.8712, 4.0481, 4.3759, 4.6717, 4.8183, 4.9126]
\tabularnewline
\midrule
2 & \seqsplit{GSPLPRPPLSPEEQEALRKKAQEKYNEFVSKIKELLRRAADRVRRGEPVELIEKTIKIGDYEYKIVATSPEEAKELENLIKEMIDLGFKPSKEFSDKLVEAARLIREGRVDEALRLLDEM} &	{'H': 61.666666666, 'B': 0.0, 'E': 11.6666666666, 'G': 0.0, 'I': 0.0, 'T': 7.5, 'S': 3.33333333333, 'P': 3.33333333333, '-': 12.5}&	[0.0207, 0.1058, 0.1782, 0.4189, 0.49, 0.9015, 1.1832, 1.8257, 2.1212, 2.8726] &	{'H': 62.5, 'B': 0.0, 'E': 11.6666666666, 'G': 0.0, 'I': 0.0, 'T': 6.6666666666, 'S': 1.66666666666, 'P': 4.1666666666, '-': 13.3333333333}&	[0.0444, 0.1641, 0.3379, 0.5724, 0.765, 0.9568, 1.4306, 1.5344, 1.6834, 1.8099]
\tabularnewline
\midrule
3 & \seqsplit{APLDPDDLSAQLRAAIDELVRLGYEEEVSKPEFIEALRLYALDLGLKEVVLRRVTPAPASQPGVYTVED VTVDLEALRKQELSPEEQARLEKIRAKYDEMLADPEFQALLDEVLARARAA} &	{'H': 57.499999999, 'B': 0.0, 'E': 13.3333333333, 'G': 0.0, 'I': 4.1666666666, 'T': 8.3333333333, 'S': 3.33333333333, 'P': 6.6666666666, '-': 6.6666666666}&	[0.7546, 1.0836, 1.5026, 1.8874, 2.0844, 2.3192, 2.7975, 3.0199, 3.0669, 3.1382]&	{'H': 61.666666666, 'B': 0.0, 'E': 15.0, 'G': 0.0, 'I': 0.0, 'T': 8.3333333333, 'S': 3.33333333333, 'P': 1.66666666666, '-': 10.0}&	[0.5256, 1.0278, 1.1566, 1.2877, 1.5521, 1.9111, 2.1887, 2.4664, 2.734, 2.8731] 
\tabularnewline
\bottomrule
\end{tabularx}
\end{table*}

\subsection{Experiment III: Protein design conditioned on the protein CATH class }\label{sec:exp_3}
CATH is a hierarchical classification system for protein structures that consists of four main levels. The highest level in this hierarchy is the “\textit{Class}” which primarily characterizes the secondary structure content of the protein. For example, $C_1$, $C_2$, and $C_3$ correspond to proteins predominantly composed of $\alpha$-helix, mainly $\beta$-sheet, and a combination of $\alpha$ and $\beta$ secondary structures. Consequently, designing proteins based on the CATH class number, i.e. $C_1$, $C_2$, $C_3$, can be understood as creating proteins with a specific fractional content of the secondary structure. Previous studies have demonstrated the importance of the protein secondary structures content, specially $\alpha$-helix/$\beta$-sheet ratio, on the mechanical properties of the protein materials\cite{Sikora2009,Nakajo2007}. For instance, $\alpha$-helix-rich proteins tend to yield stretchy materials\cite{Ackbarow2007}, while $\beta$-sheet-rich ones produce rigid materials.\cite{Knowles2011,Xu2010,Qin2010} Chroma has the potential to conditionally generate proteins with specified folds according to CATH class annotations at three levels.\cite{Ingraham2023}

\begin{figure*}
 \centering
 \includegraphics[width=\textwidth]{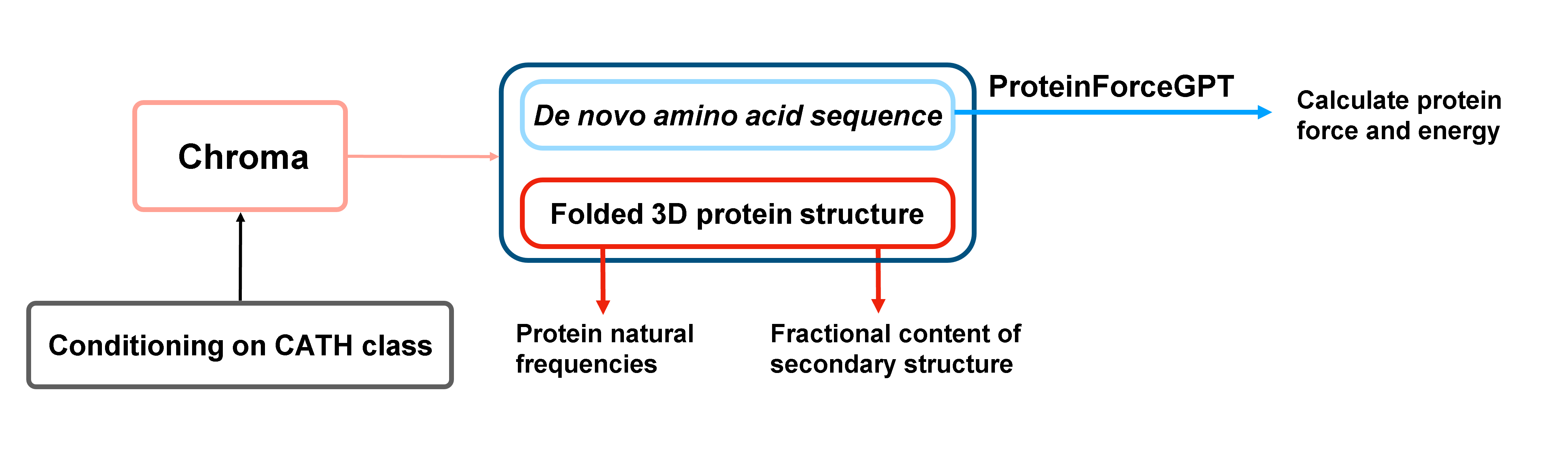}
 \caption{\textbf{Overview of the multi-agent work to solve the complex task posed in experiment III, Section \ref{sec:exp_3}.} First the multi-agent uses Chroma to generate \textit{de novo} protein sequences and structures conditioned on the input CATH class. Then using the generated protein structures, the natural frequencies and secondary structures content are computed. Next, the force (maximum force along the unfolding force-extension curve) and energy (the area under the force-extension curve) are computed from \textit{de novo} AA sequences using ProteinForceGPT.}
 \label{fig:exp3}
\end{figure*}

In this example, we task the multi-agent team with generating proteins based on their fractional content of the secondary structure and subsequently performing computational and structural analysis tasks. Specifically, in addition to secondary structure analysis and natural frequency calculations, as covered in previous examples, we instruct the team to compute the maximum unfolding force (maximum force in the unfolding force-separation curve) and unfolding energy (the area under the unfolding force-separation curve) for each generated protein. To accomplish the latter, we have equipped the multi-agent team with a custom function that utilizes a trained autoregressive transformer generative AI model, \textit{ForceGPT}. In addition to maximum unfolding force and energy, the trained generative model is able to predict the entire unfolding force-separation curve based solely on the protein amino acid sequence. Furthermore, the model has the capability to perform inverse design tasks by generating protein AA sequences that yield desired unfolding behavior. Detailed information about the training of the model can be found in \textbf{Materials and Methods} section\ref{sec:methods}. The task given is:

\begin{quote}
    \textbf{user\_proxy (to chat\_manager):\\}
    
    \textit{Design two proteins of length 100 for each family of mainly beta, mainly alpha, and alpha beta proteins. Then compute the maximum force and unfolding energy, analyze the secondary structure and compute the first 10 frequencies for these proteins.}
\end{quote}
Note that, as before, we do not specify any particular function or offer hints for selecting the appropriate function to accomplish the tasks. Instead, we empower the agents to formulate a plan, wherein they decide which functions to select and determine the input parameters. The planner outlines the following plan for the given task:

\begin{quote}
    \textbf{Planner (to chat\_manager):\\}
    
    \textit{1. Use the "design\_protein\_from\_CATH" function with inputs of length of 100, CATH\_ANNOTATION of 1 (mainly alpha), and a protein name for the first mainly alpha protein. Repeat with a different name for the second protein.\\
2. Use the same function again with inputs of length 100, CATH\_ANNOTATION of 2 (mainly beta), and a protein name for the first mainly beta protein. Repeat with a different name for the second protein.\\
3. Use the function once more with inputs of length 100, CATH\_ANNOTATION of 3 (alpha beta), and a protein name for the first alpha beta protein. Repeat with a different name for the second protein.\\
4. Use the "analyze\_protein\_structure" function on each of the protein structures obtained.\\
5. Use the "calucalte\_force\_energy\_from\_seq" function to compute the maximum force and unfolding energy for each of the proteins.\\
6. Use the "calc\_protein\_ANM" function to compute the first 10 frequencies for each protein.
}
\end{quote}
It can be seen that the planner demonstrates good performance in breaking the task into sub-tasks to be accomplished step by step. Moreover, it has identified and suggested the correct functions and corresponding input parameters for each sub-task. The plan is further supported by the “\textit{critic}” who provides positive feedback as follows:

\begin{quote}
    \textbf{Critic (to chat\_manager):\\}
    
    \textit{Yes, this plan sounds good. It covers each step required for the task. The parameter inputs for each function match with the required parameters as stated in their function definitions. Thus, this plan should successfully accomplish the task.}

\end{quote}
The multi-agent team then proceeds to execute the different steps outlined in the plan by calling and executing the functions. Specifically, the function 'design\_protein\_from\_CATH' is executed with the appropriate 'CATH\_ANNOTATION' for a specific protein structure design, as outlined in the plan. Following the generation of all proteins, the executions are followed by structural analysis and force and energy computations. It's noteworthy that the model exhibits good performance in restoring and memorizing the sequences of the generated proteins, which are essential for the force and energy calculations. Finally, the team successfully completes the task by computing the first 10 frequencies for each protein. An overview of the computations performed by the team of agents for this experiment is shown in Figure \ref{fig:exp3}.

Given the complexity of the problem involving numerous computational tasks, a decent number of results have been generated in the first round of the conversation. In the next round, to evaluate the team's ability to memorize and restore the results, we present the following task:

\begin{quote}
    \textbf{user\_proxy (to chat\_manager):\\}
    
    \textit{Could you save the results in a CSV file named "protein\_analysis.csv," where each row corresponds to a protein, and each column represents a specific property? Include the AA sequence in the results and use suitable identifiers for the columns.}

\end{quote}
In this task, we not only request the team to save the data but also require them to adhere to a customized format when storing the results. The model is proficient in creating a JSON dictionary that satisfies the specified format and saving the results to a CSV file, as illustrated in \textbf{Table \ref{tbl:exp3}}. 

The plots of the obtained results are shown in Figure \ref{fig:exp3}. The results indicate that Chroma has done a poor performance in creating $\beta$-rich protein named \textit{mainly\_beta\_protein\_2} which its structure is dominant in $\alpha$-helix. As an attempt to test the capability of the multi-agent model in analyzing the results, in the last round of the conversation, we ask the model to assess Chroma’s performance in generating the proteins conditioned on the secondary structure by posing the following question:

\begin{quote}
    \textbf{user\_proxy (to chat\_manager):\\}
    
    \textit{Based on the results of this example, can you check if the protein generator (Chroma) has been successful in creating proteins with desired structure? }

\end{quote}

The “\textit{critic}”  agent conducts a thorough evaluation of Chroma's performance in generating proteins with targeted secondary structure content. Through a detailed analysis of each CATH structure, it reveals the inherent strengths and weaknesses in Chroma's capabilities. Specifically, addressing the limitations of Chroma’s performance, the critic’s evaluation provides the following observations for the mainly beta proteins:

\begin{quote}
    \textit{- The mainly beta proteins showed higher percentages of extended strand/beta-sheet secondary structure ('E'). Though, the percentages varied quite a bit (64\% for mainly\_beta\_protein\_1 and only 8\% for mainly\_beta\_protein\_2), which could be due to the complex nature of beta-structures.}
\end{quote}

This illustration not only highlights the multi-agent model's proficiency in computational tasks but also underscores its intelligence in handling intricate data analyses—an aspect traditionally reserved for human. The full conversations for this experiment can be found in Table S4 of the Supplementary material. 

\begin{figure*}
 \centering
 \includegraphics[width=\textwidth]{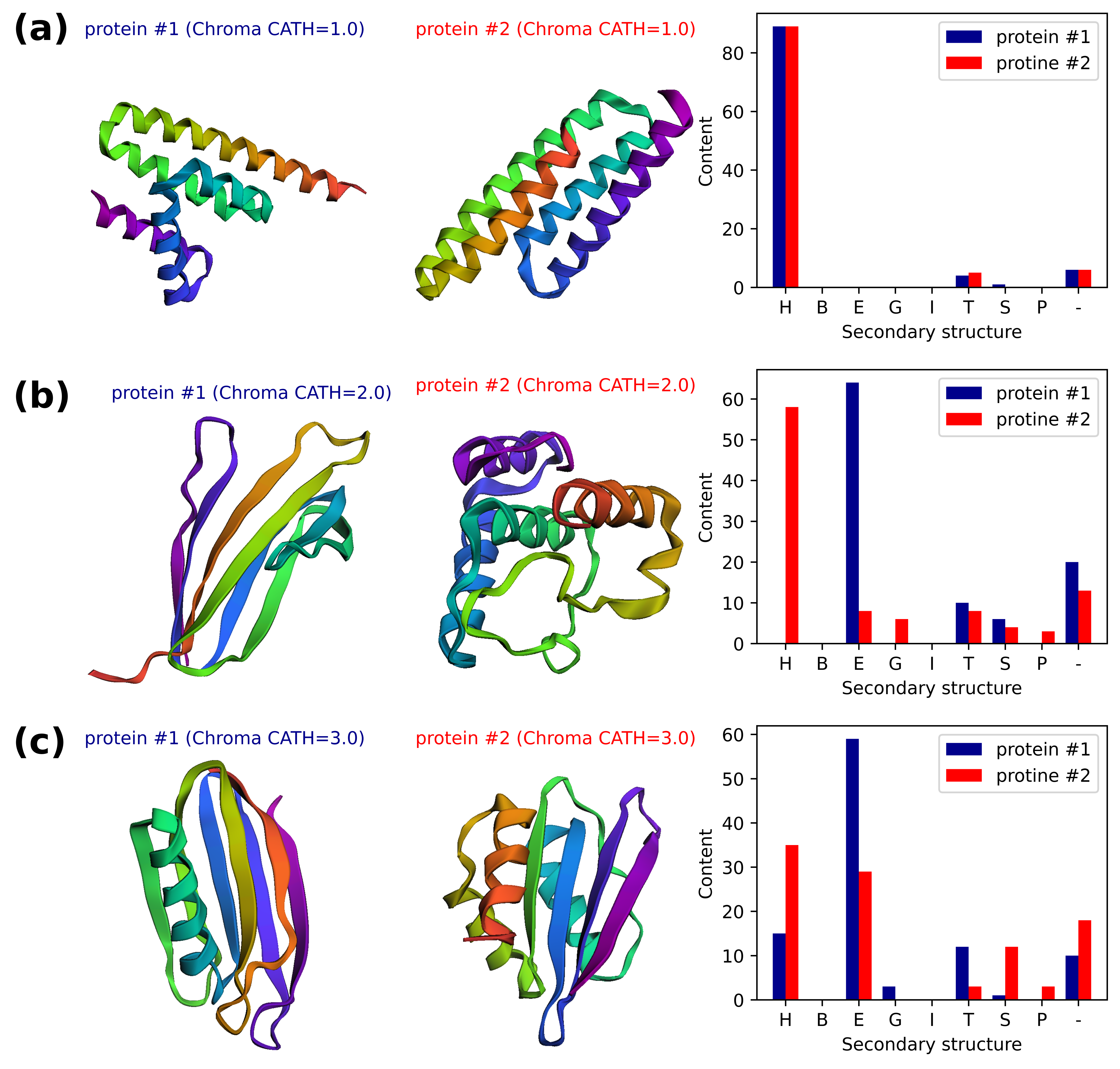}
 \caption{\textbf{The results generated by the multi-agent collaboration for the experiment III, Section \ref{sec:exp_3}.} The first and second columns depict the 3d folded structures and the last column represents the fractional content of secondary structures for the two proteins generated by Chroma conditioned on the CATH class of (a) 1: mainly alpha protein, (b) 2: mainly beta protein, and (c) 3: alpha beta protein.}
 \label{fig:chroma_CATH}
\end{figure*}

\begin{table*}
\tiny
\centering
\caption{\textbf{The final results generated by the group chat in the form of a CSV file, without human intervention, for the third experiment III, Section \ref{sec:exp_3}. } }
  \label{tbl:exp3}
\begin{tabularx}{\textwidth}{lXXrrX}
\toprule
\textbf{Protein Name\ \# } & \textbf{AA Sequence} & \textbf{Secondary Structure} & \textbf{Unfolding Energy} & \textbf{Max Force} &\textbf{First 10 Frequencies} \tabularnewline
\midrule
mainly\_alpha\_protein\_1 & \seqsplit{SMKKIEDYIREKLKALGLSDEEIEERVKQLMEGIKNPKKFE KELQKRNDRESLLIFKEAYALYEASKDKEKGKKLINKVQSERDKWETEQAEAARAAAAA} &	{'H': 89.0, 'B': 0.0, 'E': 0.0, 'G': 0.0, 'I': 0.0, 'T': 4.0, 'S': 1.0, 'P': 0.0, '-': 6.0} &	0.381	& 0.444	& \seqsplit{[0.2329, 0.4901, 0.9331, 1.3741, 1.7347, 2.1598, 2.3686, 2.6359, 2.8555, 3.0364]}
\tabularnewline
\midrule
mainly\_alpha\_protein\_2 &	\seqsplit{MSKKEIEELKKKLDEIVETLKEYARQGDDACKKAADLIEEVKKALEEGNPEKYSQLKKKLTDAINKAIEEYRKRFE AEGKPEEAQKVIDKLKKILDEITN} &	\seqsplit{{'H': 89.0, 'B': 0.0, 'E': 0.0, 'G': 0.0, 'I': 0.0, 'T': 5.0, 'S': 0.0, 'P': 0.0, '-': 6.0}} &	0.376	&0.536&	\seqsplit{[1.6126, 2.0783, 2.3073, 2.4565, 3.399, 3.475, 4.1377, 4.7104, 4.8864, 5.2187]}
\tabularnewline
\midrule
mainly\_beta\_protein\_1 &	\seqsplit{TTVTVTPPVADADGNEHSTVTAYGNKVTITITCPSNCTVTETVDGVAKTLGTVSGNQTITETRTIAPDEVVTRTYTCTPNASATSSK TQTVTIKGSQPAP} &	{'H': 0.0, 'B': 0.0, 'E': 64.0, 'G': 0.0, 'I': 0.0, 'T': 10.0, 'S': 6.0, 'P': 0.0, '-': 20.0}&	0.462&	0.533&	\seqsplit{[1.2806, 1.5057, 1.9846, 2.1025, 2.4723, 2.702, 2.9931, 3.1498, 3.4432, 4.1685]}
\tabularnewline
\midrule
mainly\_beta\_protein\_2 &	\seqsplit{SLKAKNLEEMIKEAEKLGYSRDEVEKIINEIRDKFKKLGVKISEKTLAYIAYLRLLGVKIDWDKIKKVKKATPADFRVSEEDLKKP EIQKILEKIKKEIN} &	{'H': 57.99, 'B': 0.0, 'E': 8.0, 'G': 6.0, 'I': 0.0, 'T': 8.0, 'S': 4.0, 'P': 3.0, '-': 13.0}&	0.371&	0.548&	[2.8864, 4.3752, 4.5928, 4.8295, 5.0854, 5.5618, 5.8646, 6.007, 6.3847, 7.1246]

\tabularnewline
\midrule
alpha\_beta\_protein\_1	& \seqsplit{APTVKTFEDTINGQKVTVTVTASPGGKITIKTSPGYGDEVAKAFIEELKKQNVLESYKVESAPGKETTISDVKVKSGATVTFYVINN GKKGKEYSVTVDA} &	{'H': 15.0, 'B': 0.0, 'E': 59.0, 'G': 3.0, 'I': 0.0, 'T': 12.0, 'S': 1.0, 'P': 0.0, '-': 10.0}&	0.424	&0.535&	\seqsplit{[2.4383, 2.5651, 3.3175, 3.8231, 3.9673, 4.2655, 4.6393, 5.1509, 5.6023, 5.9555]}
\tabularnewline
\midrule
alpha\_beta\_protein\_2&	\seqsplit{MELKVTEKKGKGDYKVKVIELNTPDKRYIIIESDASRESLIKAAEALLQGKEVEPTPVNEKNVVLFEDEDVKTSIERSKKLFKSDNPEE NIKKALEYLLK} &	{'H': 35.0, 'B': 0.0, 'E': 28.999999999999996, 'G': 0.0, 'I': 0.0, 'T': 3.0, 'S': 12.0, 'P': 3.0, '-': 18.0}&	0.376&	0.543	& \seqsplit{[2.8756, 3.8895, 4.0594, 4.2831, 4.5542, 5.171, 5.3661, 5.4312, 6.1964, 6.3066]}

\tabularnewline
\bottomrule
\end{tabularx}
\end{table*}

\section{Conclusions}\label{sec:conclusions}
Large Language Models (LLMs) have made remarkable strides, revealing their immense potential to potentially replicate human-like intelligence across diverse domains and modalities, demonstrating proficiency in comprehending extensive collective knowledge and proving adept at effectively applying this information. However, to reach intelligent problem-solving systems, these types of models are not yet sufficient and require integration with other methods. In this study we explored the capability of AI agents to solve protein design problems in an autonomous manner without the need for human intervention. The agents have been powered by a general purpose LLM model, GPT-4, which allows them to communicate via conversation. It should be noted that the general capabilities of the AI agents powered by the LLM plays an important role at different stages of the problem solving. In our case, GPT-4-powered agents showed excellent proficiency specially in problem understanding, strategy development, and criticizing the outcomes. Such an AI system is not limited to mere linguistic interactions between agents; they have the capacity to incorporate a variety of special-purpose modeling and simulation tools, human input, tools for knowledge retrieval, and even deep learning-based surrogate models to solve particular tasks. Furthermore, additional tools can be integrated into the multi-agent system with popular external APIs and up-to-date knowledge about special topics can be retrieved by searching and browsing the web through specialized API interfaces.  By harnessing the collective abilities of agents, including reasoning, tool usage, criticism, mutual correction, adaptation to new observations, and communication this framework has proven highly effective in navigating intricate challenges including protein design.

To achieve this goal we constructed a group of agents, each assigned a unique profile through initial prompts, to dynamically interact in a group chat via conversations and make decisions and take actions based on their observations. The agents profile outlines their attributes, roles, and functionalities within the system and describe communication protocols to exchange information with other agents in the system. Our team of agents include a user proxy to pose the query, a planner to formulate a plan, an function-baked assistant to execute the functions, and a critic to evaluate the outcome and criticizing the performance. We also use a chat manager to lead the group chat by dynamically choosing the working agent based on the current outcome and the agents' roles. Through a series of experiments, we unleashed the power of agents in not only conducting the roles they were assigned to, but to autonomously collaborate by discussion powered by the all-purpose LLM. For example, the agent playing the role of a planner successfully identified all the tasks in the query and suggested a details plan including the necessary functions to accomplish them. Furthermore, the agent assigned the critic role, is able to give constructive feedback about the plan or provide suggestions in case of failure, to correct errors that may emerge. Our experiments have showcased the great potential of the multi-agent modeling framework in tackling complex tasks as well as integrating AI-agents into physics-based modeling. 

Multi-agent modeling is a powerful technique that offers enhanced problem-solving capacity as shown here in various computational experiments in the realm of protein design, physics modeling, and analysis. 
Given a complex query comprising multi-objective tasks, using the idea of division of labor, the model excels at developing a strategy to break the task into sub-tasks and then, recruiting a set of agents to effectively engage in problem solving tasks in an autonomous fashion. Tool-backed agents have the capacity to execute tools via function execution. We equipped an agent with a rich library of tools that span a broad spectrum of functionalities including \textit{de novo} protein design, protein folding, and protein secondary structure analysis among others. The fact that there is no intrinsic limitation in customizing the functions, allows us to integrate knowledge across different disciplines into our model and analysis, for instance by integrating knowledge retrieval systems or retrieving physical data via simulations. For instance, here we utilized coarse grained simulations to obtain natural frequencies of proteins but the model offers a high flexibility in defining functions that focus on other particular area simulation (e.g. an expert in performing Density Functional Theory, Molecular Dynamics, or even physics-inspired neural network solvers\cite{Lagaris1998,Raissi2019,Peng2021}). Multi-agent framework can also accelerate the discovery of \textit{de novo} proteins with targeted mechanical properties by embracing the power of robust end-to-end deep models solving forward and inverse protein design problems\cite{Hu2022,Lu2023kaplan,Yu2022,Liu2022,Yu2020,Buehler2023autoregressive,Ni2023secondary}

Developing these models that connect some structural protein features, such as secondary structure, to a material property, such as toughness or strength have gained a lot of attention recently. Here, we used a pre-trained autoregressive transformer model to predict the maximum force and energy of protein unfolding, but other end-to-end models could also be utilized.  In the context of inverse protein design problems, a team of two agents, one expert in the forward tasks and the other in the inverse task, can be collaborated to assist the cycle check wherein the \textit{de novo} proteins certainly meet the specified property criteria. Along the same line, one could benefit from the multi-agent collaboration in evaluating the accuracy of generative models in conditional designing of proteins or compare the created 3D structures with the state-of-the art folding tools\cite{Jumper2021,AlQuraishi2021,Gao2020}. For example, through an automated process of protein generation and structure analysis, our ProtAgents framework revealed the shortcomings of Chroma in designing $\beta$-sheet-rich proteins. In another example, the folded 3D structures of Chroma were compared with those obtained by OmegaFold2. All these examples, demonstrate the capacity of multi-agent framework in a wide range of applications in the context of protein design and analysis.  Lastly, the model enables integrating various information across scales, whether new protein sequences or physics simulations output in form of rich data structures, for inclusion in easily readable file formats (like JSON) to be used by other agents or to be stored for future analysis. 

Designing \textit{de novo} proteins that meet special objectives in term of mechanical or structural properties present unique challenges calling for new strategies. The prevailing strategies often rely on developing data-driven end-to-end deep learning models to find the complex mapping from protein constitutive structure to property or vice versa. However, these models often focus on specific properties, limiting their functionality in multi-objective design purposes where several criteria needs to be met. To overcome these challenges and propel the field forward, future research endeavors could revolve around the development of an integrated system of agents designed to automate the entire lifecycle of training deep neural networks for protein design. Each agent within this system could be assigned specific responsibilities, such as data generation through simulations, data curation for ensuring quality and relevance, and the execution of the code required for model training. Additionally, a critic agent could monitor and critique the training process, making decisions like early stopping or tuning hyperparameters to enhance the model's accuracy. This collaborative and automated approach would not only streamline the design process but also contribute to achieving higher or desired levels of accuracy in the generated models. Furthermore, this agent-based strategy can extend to on-the-fly active learning, where agents dynamically adapt the model based on real-time feedback, improving its performance iteratively. By incorporating intelligent agents at every stage of the process, the proposed system aims to revolutionize the landscape of \textit{de novo} protein design, making it more efficient, adaptive, and capable of meeting diverse and complex design objectives, offering a new paradigm in materials design workflows.

\section{Materials and Methods}\label{sec:methods}

\subsection*{Agent design}

As shown in Figure \ref{fig:agents}a, we design AI agents using the state-of-the-art all-purpose LLM GPT-4 and dynamic multi-agent collaboration is implemented in AutoGen framework\cite{Zhu2023}, an open-source ecosystem for agent-based AI modeling. Additional agents are introduced as described below, including some based on generative AI as well as physics modeling. 

In our multi-agent system, the human \textit{user\_proxy} agent is constructed using UserProxyAgent class from Autogen, and \textit{Assistant}, \textit{Planner}, \textit{Critic} agents are created via AssistantAgent class from Autogen; and the group chat manager is created using GroupChatManager class. Each agent is assigned a role through a profile description listed in \textbf{Table \ref{tbl:agents_profile}}, included as \textit{system\_message} at their creation.   

\subsection*{Function and tool design}
All the tools implemented in this work are defined as python functions. Each function is characterized by a name, a description, and input properties with a description as tabulated in \textbf{Table S1} of the Supplementary Material. The list of functions are incorporated into the multi-agent system, included as the \textit{function\_map} parameter in the Assistant agent at its creation. 

\subsection*{Autoregressive transformer model to predict protein unfolding force-extension from sequences}

We use a special-purpose GPT-style model denoted as ProteinForceGPT, similar as  in , here trained to predict force-extension curves from sequences along with other mechanical properties, and vice versa (\url{https://huggingface.co/lamm-mit/ProteinForceGPT}). The protein language model is based on the NeoGPT-X architecture and uses rotary positional embeddings (RoPE)\cite{su2024roformer}. The model has 16 attention heads, 36 hidden layers and a hidden size of 1024, an intermediate size of 4096 and uses GeLU activation functions. 

Pre-training was conducted based on a dataset of $\sim800,000$ amino acid sequences, using next-token predictions using a “Sequence” task 
(\url{https://huggingface.co/datasets/lamm-mit/GPTProteinPretrained}):
\begin{quote}
\tiny {
\texttt{
Sequence<GEECDCGSPSNPCCDAATCKLRPGAQCADGLCCDQCRFKKKRTICRIARGDFPDDRCTGQSADCPRWN>
}
}
\end{quote}

The ProteinForceGPT model was then fine-tuned bidirectionally, to predict mechanical properties of proteins from their sequence, as well as sequence candidates that meet a required force-extension behavior and various other properties. Fine-tuning is conducted using a dataset derived from molecular dynamics (MD) simulations\cite{Ni2023ForceGen}. Sample tasks for the model include:

\begin{quote}
\tiny {\texttt{CalculateForce<GEECDCGSPSNPCCDAATCKLRPGAQCADGLCCDQCRFKKKRTICRIARGDFPDDRCTGQSADCPRWN> [0.262]\\
CalculateEnergy<GEECDCGSPSNPCCDAATCKLRPGAQCADGLCCDQCRFKKKRTICRIARGDFPDDRCTGQSADCPRWN> [0.220]\\ CalculateForceEnergy<GEECDCGSPSNPCCDAATCKLRPGAQCADGLCCDQCRFKKKRTICRIARGDFPDDRCTGQSADCPRWN>\,[0.262,0.220]\\
CalculateForceHistory<GEECDCGSPSNPCCDAATCKLRPGAQCADGLCCDQCRFKKKRTICRIARGDFPDDRCTGQSADCPRWN>  
 [0.004,0.034,0.125,0.142,0.159,0.102,0.079,0.073, 0.131,0.105,0.071,0.058,0.072,0.060,0.049,0.114, 0.122,0.108,0.173,0.192,0.208,0.153,0.212,0.222,
 0.244]\\
GenerateForce<0.262>\,[GEECDCGSPSNPCCDAATCKLRPGAQCADGLCCDQCRFKKKRTICRIARGDFPDDRCTGQSADCPRWN]\\
GenerateForce<0.220>\,[GEECDCGSPSNPCCDAATCKL RPGAQCADGLCCDQCRFKKKRTICRIARGDFPDDRCTGQSADCPRWN]\\
GenerateForceEnergy<0.262,0.220>\,[GEECDCGSPSNPCCDAATCKLRPGAQCADGLCCDQCRFKKKRTICRIARGDFPDDRCT\,
GQSADCPRWN]\\
GenerateForceHistory<0.004,0.034,0.125,0.142,0.159, 0.102,0.079,0.073,0.131,0.105,0.071,0.058,0.072,
0.060, 0.049,0.114,0.122,0.108,0.173,0.192,0.208,
0.153,0.212,\,
0.222,0.244>
[GEECDCGSPSNPCCDAATCKLR
PGAQCADGLCCDQCRFKKKRTICRIARGDFPDDRCTGQSADCPRWN]
}}
\end{quote}

Sample results from validation of the model are shown in \textbf{Figure S2}. We only use forward predictions for use in the agent model reported here. 

\subsection*{Software versions and hardware}
We develop our multi-agent models using local workstations with NVIDIA GPUs. We use Python 3.10 and pyautogen-0.2.2\cite{Wu2023}. Additional implementation details are included in the code. 

\subsection*{Visualization}
We use Py3DMol\cite{Rego2015} for visualization of the protein structures. 

\subsection*{Secondary Structure Analysis}
We use the dictionary of protein secondary structure (DSSP)\cite{Kabsch1983DictionaryFeatures} module via BioPython\cite{Cock2009} to analyze the secondary structure content of the proteins from its geometry.

\subsection*{Natural Vibrational Frequency Calculations}
We perform Anisotropic Network Model (ANM)\cite{Atilgan2001,Doruker2000} calculations as implemented in ProDy\cite{Bakan2011} for normal mode analysis. The problem is solved by considering the protein as a network of interactions, defined within a cutoff distance for which spring-like potentials are assumed to define molecular interactions. 

\subsection*{Retrieval Augmented Generation}
We use Llama Index\cite{llama_index} as a tool to implement RAG where the full text of papers cited as references \cite{Sikora2009,Nakajo2007} are used as external sources from which information can be retrieved by the system in real-time. 

\subsection*{Conflict of interest}
The author declares no conflict of interest.

\subsection*{Data and code availability}
All data and codes are available on GitHub at \url{https://github.com/lamm-mit/ProtAgents}. Alternatively, they will be provided by the corresponding author based on reasonable request.  
\\
\\
\textbf{Author Contributions:}
MJB and AG conceived the study and developed the multi-agent models. AG performed the tests for various problems, analyzed the results and prepared the first draft of the paper. MJB supported the analysis, revised and finalized the paper with AG. 

\subsection*{Supplementary Materials}
The full records of different conversation experiments along with additional materials are provided as Supplementary Materials. 

\section*{Acknowledgements}
We acknowledge support from USDA (2021-69012-35978), DOE-SERDP (WP22-S1-3475), ARO (79058LSCSB, W911NF-22-2-0213 and W911NF2120130) as well as the MIT-IBM Watson AI Lab, MIT’s Generative AI Initiative, and Google. Additional support from NIH (U01EB014976 and R01AR077793) ONR (N00014-19-1-2375 and N00014-20-1-2189) is acknowledged. AG gratefully acknowledges the financial support from the Swiss National Science Foundation (\#P500PT\_214448).

%Bibliography

\bibliographystyle{unsrt}  
\bibliography{library,references} %You need to replace "rsc" on this line with the name of your .bib file

\end{document}